\documentclass{JAC2003}

\usepackage{graphicx}

\begin{document}

\title{KLOE PERSPECTIVES FOR R-MEASUREMENTS AT DAFNE2}

\author{ The KLOE collaboration\thanks{
The KLOE Collaboration: A.~Aloisio,
F.~Ambrosino,
A.~Antonelli,
M.~Antonelli,
C.~Bacci,
G.~Bencivenni,
S.~Bertolucci,
C.~Bini,
C.~Bloise,
V.~Bocci,
F.~Bossi,
P.~Branchini,
S.~A.~Bulychjov,
R.~Caloi,
P.~Campana,
G.~Capon,
T.~Capussela,
G.~Carboni,
G.~Cataldi,
F.~Ceradini,
F.~Cervelli,
F.~Cevenini,
G.~Chiefari,
P.~Ciambrone,
S.~Conetti,
E.~De~Lucia,
P.~De~Simone,
G.~De~Zorzi,
S.~Dell'Agnello,
A.~Denig,
A.~Di~Domenico,
C.~Di~Donato,
S.~Di~Falco,
B.~Di~Micco,
A.~Doria,
M.~Dreucci,
O.~Erriquez,
A.~Farilla,
G.~Felici,
A.~Ferrari,
M.~L.~Ferrer,
G.~Finocchiaro,
C.~Forti,
A.~Franceschi,
P.~Franzini,
C.~Gatti,
P.~Gauzzi,
S.~Giovannella,
E.~Gorini,
E.~Graziani,
M.~Incagli,
W.~Kluge,
V.~Kulikov,
F.~Lacava,
G.~Lanfranchi,
J.~Lee-Franzini,
D.~Leone,
F.~Lu,
M.~Martemianov,
M.~Matsyuk,
W.~Mei,
L.~Merola,
R.~Messi,
S.~Miscetti,
M.~Moulson,
S.~M\"uller,
F.~Murtas,
M.~Napolitano,
A.~Nedosekin,
F.~Nguyen,
M.~Palutan,
E.~Pasqualucci,
L.~Passalacqua,
A.~Passeri,
V.~Patera,
F.~Perfetto,
E.~Petrolo,
L.~Pontecorvo,
M.~Primavera,
F.~Ruggieri,
P.~Santangelo,
E.~Santovetti,
G.~Saracino,
R.~D.~Schamberger,
B.~Sciascia,
A.~Sciubba,
F.~Scuri,
I.~Sfiligoi,
A.~Sibidanov,
T.~Spadaro,
E.~Spiriti,
M.~Testa,
L.~Tortora,
P.~Valente,
B.~Valeriani,
G.~Venanzoni,
S.~Veneziano,
A.~Ventura,
S.~Ventura,
R.~Versaci,
I.~Villella,
G.~Xu.
} presented by Achim G. Denig\thanks{Invited talk given at the workshop '$e^+e^-$ in the 1-2 GeV range: 
Physics and Accelerator Prospects', Alghero/Sardinia, 
Sept.10-13,2003,
Internet: http://www.lnf.infn.it/conference/d2/gener.html
}\\
  {\small Universit\"at Karlsruhe, IEKP, Postfach 3640, 76021 Karlsruhe, Germany}}

\maketitle

\begin{abstract}
As a future upgrade of the Frascati $\phi$ factory
DA$\Phi$NE an increase of the center-of-mass energy of the 
accelerator up to $W=2$ GeV has been proposed (DAFNE2). 
In this case the hadronic cross section in the energy range
between $1-2$ GeV can be measured with the KLOE detector. 
The feasibility of these measurements and the impact 
on the hadronic contribution to the anomalous magnetic moment of
the muon, $a_{\mu}^{\rm hadr}$, are discussed. The possibilities for
an energy scan are compared with the radiative return technique, 
in which the accelerator is running at a fixed center-of-mass energy
and ISR-events are taken to lower the invariant mass of the 
hadronic system. 
\end{abstract}

\section{The DAFNE2 Proposal}
The future perspectives of the $e^+e^-$ collider 
DA$\Phi$NE are discussed at the Frascati laboratories 
(see also~\cite{dafne}\cite{superdafne}).
Two projects have been proposed recently: an increase of the peak luminosity to 
$\approx 10^{34}cm^{-2}s^{-1}$ (DA$\Phi$NE-II) and an increase of the 
center-of-mass energy up to $2 {\rm GeV}$ (DAFNE2, \cite{dafne2}), 
where the second option 
might be realized either before or within the high-luminosity-solution. 
While at DA$\Phi$NE-II the main physics motivation is based on the investigation of the 
parameters of the kaon system (CP,CPT-violation), 
DAFNE2 provides the possiblity to measure
the timelike nucleon form factors at threshold and to perform hadronic cross section
measurements in the $1-2 {\rm GeV}$ energy range, which we will discuss in
the following.
Some of the components of the present machine are already  
designed for an energy increase. The main hardware
modifications concern the dipole magnets, the splitter
magnets and the low-$\beta$ quadrupoles.
No crucial issues from the accelerator physics point of view can
be seen at the moment. Peak luminosities of at least $10^{32}{\rm cm}^{-2}{\rm s}^{-1}$
are expected to be in reach for this machine, which allows to collect an integrated
luminsity per year of at least $1$fb$^{-1}$.

\section{Importance of the 1-2 GeV energy range}
Hadronic cross section data are of importance for the determination
of the hadronic contribution to the anomalous magnetic moment of the 
muon, $a_{\mu}$, and for the fine structure constant at the $Z$ pole,
$\alpha(m_Z^2)$. In the following we will discuss the impact of
DAFNE2 on the muon anomaly. The hadronic contribution to 
this fundamental quantity, $a_\mu^{\rm hadr}$, which is given by
the hadronic vacuum polarization, cannot be calculated at low energies using perturbative QCD.
A dispersion relation can however be derived,
giving $a_\mu^{\rm hadr}$ as an integral over the hadronic 
cross section, multiplied by an appropriate kernel. 
The dominant contribution to $a_{\mu}^{\rm hadr}$  ($90\%$) is given
by low energy cross section measurements $<2{\rm GeV}$~\cite{davhoe2}. 
This is the region where DAFNE2 will operate. An improved measurement
of hadronic cross sections for the various channels of interest could therefore 
considerably improve the knowledge on the hadronic contribution to $a_{\mu}$.
This is 
needed for an interpretation of the recent new 
measurements \cite{e821a}\cite{e821b}
of the muon anomaly
(E821 collaboration, BNL), showing a difference
between the experimental and theoretical value of $a_\mu$ of up to $3\sigma$
(see ref.~\cite{davhoe2} for details concerning the theory evaluation).
\\
In table~\ref{tab:tbl2} the contributions to $a_{\mu}^{\rm hadr}$ 
and to the squared error $\delta^2 a_{\mu}^{\rm hadr}$ are listed for
different energy ranges and for different hadronic channels~\footnote{
Ref.~\cite{davhoe2} has been used 
for this calculation}.  It is interesting to notice that the $2\pi$ channel is contributing
to $a_{\mu}^{\rm hadr}$ to $54\%$ around the $\rho$ peak ($0.6 - 1.0 {\rm GeV}$),
while the contribution to the error $\delta^2 a_{\mu}^{\rm hadr}$ in the 
same energy interval is only $34\%$. 
This difference reflects the fact that the $2\pi$ channel around the $\rho$ peak  
is very well measured now by CMD-2~\cite{cmd2}~\cite{cmd2b} with a systematic error of $0.6\%$.
Soon also KLOE will publish its results of this channel and of 
the same energy interval (see these proceedings ref. \cite{smueller}). Precision measurements for
this channel exist also 
from the analysis of hadronic $\tau$ decays which are related via the CVC-theorem to
electron-positron data and can be used for the evaluation of $a_{\mu}^{\rm hadr}$ 
after appropriate isospin corrections.  
At low energies ($<0.6 {\rm GeV}$) and even more important at high energies ($>1 {\rm GeV}$) 
a considerable improvement for the two-pion channel is required. 
The contribution to the error $>1 {\rm GeV}$ is 
very large ($\approx 30\%$) while the absolute contribution to the integral $a_{\mu}^{\rm hadr}$ 
is rather small (only $10\%$). DAFNE2 can play an important role here.
The threshold region ($<0.6$ GeV) will be measured already at the present DA$\Phi$NE
machine in a complementary analysis to the one published now.
\\
Another interesting hadronic channel is the $4\pi$ channel where measurements
with a precision not better than $10-20\%$ exist. The $4\pi$ channel becomes important
only above $1 {\rm GeV}$ and is therefore a good candidate for DAFNE2. 
This will be discussed in more detail in the following. The relative
contribution of the $4\pi$-channel to the error $\delta^2 a_{\mu}^{\rm hadr}$ 
is $7\%$.

\begin{figure}
\includegraphics[height=0.5\textwidth]{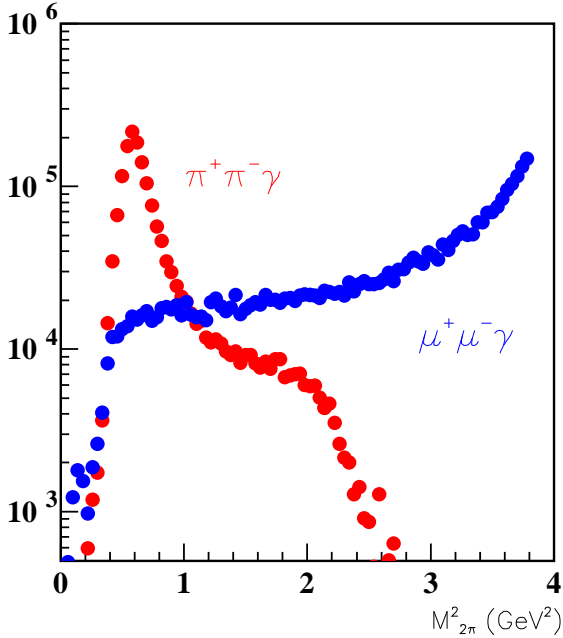}
\caption{$\pi^+\pi^-\gamma$ event yield for an integrated luminosity of $1 {\rm fb}^{-1}$.
Realistic acceptance cuts have been applied: $\Theta_{\pi}>30^o$,  $\Theta_{\gamma}<20^o$
or $\Theta_{\gamma}>160^o$. The yield for radiative muon pair production is also shown. 
The statistics is sufficient to normalize the cross section 
measurement to muon pairs.}
\label{fig:fig2}
\end{figure}

\begin{figure}
\includegraphics[height=0.5\textwidth]{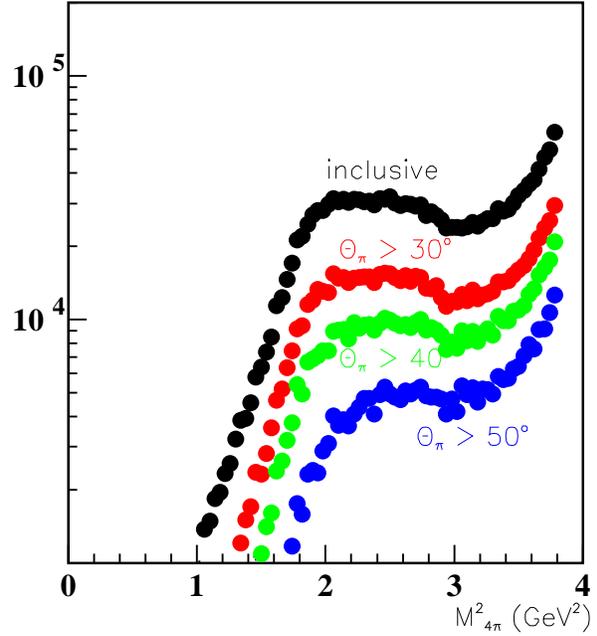}
\caption{$\pi^+\pi^-\pi^+\pi^-\gamma$ event yield for an integrated luminosity of $1 {\rm fb}^{-1}$.
The dependence of the event rate with the polar angle cut for the pion tracks is shown. The 
radiated photon is selected at small angles in order to decrease the relative contribution of
FSR events: $\Theta_{\gamma}<20^o$ or $\Theta_{\gamma}>160^o$}
\label{fig:fig3}
\end{figure}

\section{Energy Scan versus Radiative Return}
Up to recently an energy scan has been considered as the only way to 
measure hadronic cross sections $e^+e^-\to {\rm hadrons}$ at electron-positron colliders. 
The KLOE and BABAR \cite{md_babar} collaborations
have shown in the meanwhile that the use of Initial State Radiation (ISR) events 
has to be considered as a complementary and competitive approach at particle
factories, which are actually designed for fixed center-of-mass energies W.
In this new  method - called also 'radiative return' - hadronic 
events are taken, in which a photon (energy $E_\gamma$) is radiated before annihilation of the 
$e^+e^-$ pair. The invariant mass $M^2_{\rm hadr}$ of the hadronic system is given by: 
$M^2_{\rm hadr}=W^2-2WE_\gamma$. 
In general the cross sections  
$\sigma_{{\rm hadr}+\gamma}=\sigma(e^+e^-\to{\rm hadrons}+\gamma)$
and $\sigma_{\rm hadr}=\sigma(e^+e^-\to {\rm hadrons})$ are related through:
\begin{equation}
M^2_{\rm hadr} \cdot \frac{d\sigma_{{\rm hadr}+\gamma}}{dM^2_{\rm hadr}}=
\sigma_{\rm hadr} \cdot H(M^2_{\rm hadr})
\label{eq:H}
\end{equation}
The radiator function $H$ is taken from theory. In the case of KLOE we use the
PHOKHARA generator, designed specially for our purposes (see below).

\begin{center}
\begin{table*}[htb]
\caption{Relative contributions of different hadronic channels and energy ranges to 
$a_{\mu}^{\rm hadr}$ and $\delta^2 a_{\mu}^{\rm hadr}$~\cite{davhoe2}.}
\label{table:2}
\newcommand{\m}{\hphantom{$-$}}
\newcommand{\cc}[1]{\multicolumn{1}{c}{#1}}
\renewcommand{\tabcolsep}{2pc}    
\renewcommand{\arraystretch}{1.2} 
\begin{tabular}{llll}
\hline
Channel & Energy Range& $a_\mu^{\rm hadr}$ (rel. contr.) & $\delta^2 a_{\mu}^{\rm hadr}$ (rel. contr.) \\
\hline
$2\pi$ & $2m_{\pi}-0.5{\rm GeV}$ &  $8\%$ &  $8\%$  \\
$2\pi$ & $0.6-1.0{\rm GeV}$      & $54\%$ & $34\%$  \\
$2\pi$ & Rest $<1.8{\rm GeV}$    & $10\%$ & $31\%$  \\
$3\pi$ & $<1.8{\rm GeV}$         & $12\%$ &  $5\%$  \\
$4\pi$ & $<1.8{\rm GeV}$         &  $5\%$ &  $7\%$  \\
$>4\pi$ & $<1.8{\rm GeV}$        &  $3\%$ &  $5\%$  \\
\hline
\label{tab:tbl2}
\end{tabular}\\[2pt]
\end{table*}
\end{center}

The design of the DAFNE2 project foresees the possibility of a systematic variation of the
center-of-mass energy in the $1-2$ GeV range~\footnote{and the possibility to measure the
center-of-mass of the machine very precisely using the resonant depolarization technique}.
An energy scan is thus possible. 
However, also the radiative return is an option for DAFNE2
when the center-of-mass energy of the accelerator is kept fixed at e.g. $W=2$ GeV
or close to the $N\bar{N}$ threshold region where measurements of the timelike
nucleon form factor are planned.
In the following we will briefly point out the advantages and possible issues of
the radiative return method compared to an energy scan.
\\
One big advantage of the method is the fact that data comes as
a by-product of the standard program of the machine (e.g. CP violation measurements 
in the case of KLOE/BABAR) and no dedicated experimental
modifications are needed. Moreover, the method allows to measure the whole energy 
spectrum below the center-of-mass energy of the accelerator at a time. 
Systematic errors from luminosity, the knowledge of the 
machine energy, efficiencies and acceptances have to be determined only 
for one single energy point (as a function of $M^2_{\rm hadr}$ though) and not for each energy
bin as it is needed in the case of an energy scan.
\\
There are on the other side a series of issues which need to be attacked, especially if the
radiative return method is used for a high precision measurement on the level of
$1\%$ or below. Clearly the method requires a precise theoretical know\-ledge 
of the ISR-process, i.e. of the radiator function $H$ in equation 1. 
A lot of progress has been obtained in the last years and
calculations exist now up to NLO by means of the 
Monte Carlo generator PHOKHARA~\cite{binner}~\cite{german}~\cite{czyz}
~\cite{grzelinska}~\cite{grzelinska2}. 
Another important issue is the suppression of FSR events, since FSR has to be
considered as a background to the ISR-approach of the radiative 
return \cite{denig99}. 
Unfortunately, the radiation of photons from hadrons can only be 
calculated within a certain model dependence. Usually the model of
scalar QED is chosen for the radiation of photons from e.g. pions.
The actual KLOE analysis uses events in which the radiative photon is selected at
small angles, which effectively suppresses the relative amount of FSR well below 
$1\%$, such that the model dependence becomes negligible.  
Moreover, the validity of the model for FSR can be tested from data by 
measuring the charge asymmetry~\cite{binner}~\cite{slac}~\cite{hoefer} and by
comparing the model prediction with data.

\begin{figure}
\includegraphics[height=0.5\textwidth]{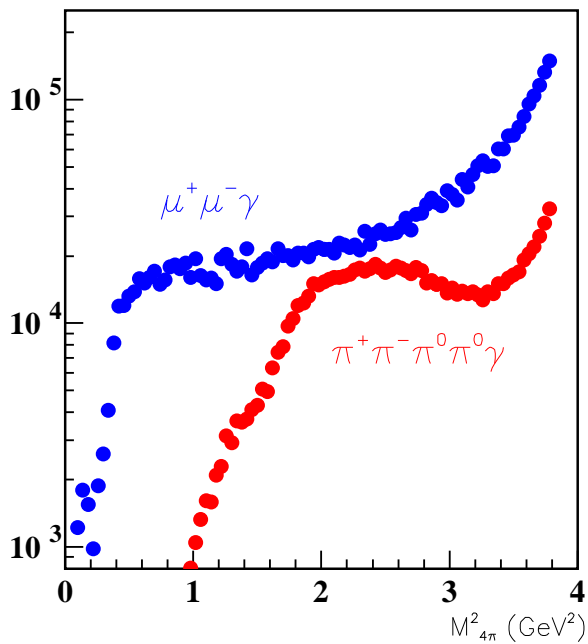}
\caption{$\pi^+\pi^-\pi^0\pi^0\gamma$ event yield for an integrated luminosity of $1 {\rm fb}^{-1}$
and for realistic selcetion cuts (see figure caption 1). The number of $\mu^+\mu^-\gamma$ events
is again shown for comparison. 
}
\label{fig:fig4}
\end{figure}

\section{Radiative Return at DAFNE2}
The PHOKHARA Monte Carlo code has been used to study the event rates 
for ISR-events at DAFNE2. The machine is assumed to operate at
$\sqrt{s}=2 {\rm GeV}$. We have investigated the two-pion-state $\pi^+\pi^-$ 
and the four-pion-states $\pi^+\pi^-\pi^+\pi^-$ and $\pi^+\pi^-\pi^0\pi^0$
in the $1-2 {\rm GeV}$ range due to their special importance for  
the hadronic contribution to the muon anomlay (see above).
In figures~\ref{fig:fig2}, \ref{fig:fig3} and~\ref{fig:fig4} the $M^2_{\rm hadr}$ 
differential event rates for the
states $\pi^+\pi^-\gamma$, $\pi^+\pi^-\pi^+\pi^-\gamma$ 
and $\pi^+\pi^-\pi^0\pi^0\gamma$ are shown, where $M^2_{\rm hadr}$ is the 
invariant mass of the hadronic (muonic) system. A bin width of $0.04 {\rm GeV}$ 
has been chosen. The plots show the event yield for an integrated 
luminosity of $1 fb^{-1}$ and for realistic acceptance cuts (see figure caption
for more details). There are no limitations from statistics, since
the event yield is in the order $10.000$ events in almost the entire energy range of interest.
In fig.~\ref{fig:fig2} and~\ref{fig:fig4} 
in addition to the hadronic channels the yield of $\mu^+\mu^-\gamma$ events is shown, 
proving that a normalization to muon events is feasible from the statistical 
point of view. In the following we briefly present the main experimental issues to be studied:

\begin{itemize}

\item{The KLOE drift chamber \cite{dc} allows a high resolution measurement
of the invariant mass $M^2_{\rm hadr}$ for the fully charged hadronic channels.
In the case of the $\pi^+\pi^-\pi^0\pi^0\gamma$ channel the experimental challenge is
the correct $\pi^0$ reconstruction and a possible unfolding of the mass spectrum due to the
limited resolution of the KLOE electromagneic calorimeter \cite{emc}.}

\item{The suppression of FSR is of great importance for a successful application of 
the radiative return (see discussion
above). Fortunately at $W=2 {\rm GeV}$ the pion form factor is very small such that the
relative amount of FSR in the two-pion-channel will be reduced also.}

\item{In contrary to the present KLOE analysis there will be no background from $\phi$ decays
(e.g. $\phi \to \pi^+\pi^-\pi^0$) and therefore a much reduced background contamination
can be expected at DAFNE2. Moreover, also the Bhabha cross section is considerably
reduced with respect to the present DA$\Phi$NE machine.}

\item{Above $M^2_{\rm hadr}=2$ GeV$^2$ the two-pion cross section is decreasing rapidly 
(see fig.~\ref{fig:fig2}) while the muonic cross section is high. An efficient separation 
of pions and muons might become critical in this region.}

\item{KLOE does not have experience in the measurements of channels where four tracks
are originating from the interaction point. Special reconstruction software has to be 
developed for the analysis of the $\pi^+\pi^-\pi^+\pi^-\gamma$ channel.}

\end{itemize}

In order to understand the final precision for these radiative return measurements,
a dedicated feasibility study, including the KLOE detector simulation environment,
is needed. We want to stress that no a-priori limitations for a measurement
on the level of few percent can be seen at the moment. 
This is sufficient for a sizeable reduction of the contribution above 
$1$ GeV to the error on $a_\mu^{\rm hadr}$.
The experimental issues discussed above, are similar in the case of an 
energy scan and do not represent a drawback of the radiative return method.

\section{Conclusions}
DAFNE2 provides the possiblity to measure the
hadronic cross section in the $1-2 {\rm GeV}$ energy range. 
The radiative return seems a feasible option for these cross section measurements.
Special emphasis should be put on the two-pion and four-pion-channels $>1{\rm GeV}$
due to their importance for an  
improved evaluation of the hadronic contribution to the muon anomaly. The long
term goal is a reduction of the error of the hadronic contribution to the muon
anomaly to a value $\delta a_{\mu}^{\rm hadr}=2\cdot10^{\rm -10}$. 
DAFNE2 can make a considerable contribution to this goal. Competition comes
from the radiative return activities at BABAR and possible
future activities at VEPP-2000, BELLE and CLEO-c.

\end{document}